\documentclass{bmcart}

\usepackage{amsthm,amsmath}
\usepackage{braket}
\usepackage[utf8]{inputenc} 
\usepackage{hyperref}
\usepackage{indentfirst}

\usepackage{multicol}
\usepackage{array}
\usepackage{graphicx}
\usepackage{cancel}
\usepackage{braket}
\usepackage{caption}
\usepackage{subcaption}
\usepackage{float}
\usepackage{amssymb}
\usepackage{bbold}
\usepackage{slashed}
\usepackage{color}
\usepackage{cite}
\usepackage{siunitx}
\startlocaldefs

\endlocaldefs

\begin{document}

\begin{frontmatter}

\begin{fmbox}
\dochead{Research}


\title{Digital quantum simulation of gravitational optomechanics with IBM quantum computers}

\author[
  addressref={aff1},
  email={}
]{\inits{P.G.}\fnm{Pablo Guillermo} \snm{Carmona Rufo}}
\author[
  addressref={aff2},
  email={}
]{\inits{A.}\fnm{Anupam} \snm{Mazumdar}}
\author[
  addressref={aff3},
  email={}
]{\inits{S.}\fnm{Sougato} \snm{Bose}}
\author[
  addressref={aff4},                   
  email={carlos.sabin@uam.es}   
]{\inits{C.}\fnm{Carlos} \snm{Sabín}}
%

\address[id=aff1]{%
  \orgdiv{Instituto de Física Teórica},
\orgname{UAM/CSIC},
  \street{C/ Nicolás Cabrera 13-15, Campus de Cantoblanco},
  \postcode{28049}
  \city{Madrid},
  \cny{Spain}
}

\address[id=aff2]{%
  \orgdiv{Van Swinderen Institute for Particle Physics and Gravity},
\orgname{University of Groningen},
  \street{},
  \postcode{9747AG}
  \city{Groningen},
  \cny{the Netherlands}
}

\address[id=aff3]{%
  \orgdiv{Department of Physics and Astronomy},
\orgname{University College London},
  \street{},
  \postcode{WC1E 6BT}
  \city{London},
  \cny{United Kingdom}
}

 \address[id=aff4]{
  \orgdiv{Departamento de Física Teórica},             
  \orgname{Universidad Autónoma de Madrid},          
 \postcode{28049},
  \city{Madrid},              
  \cny{Spain}                 
}



\end{fmbox}


\begin{abstractbox}

\begin{abstract} 

We showcase the digital quantum simulation of the action of a Hamiltonian that governs the interaction between a quantum mechanical oscillator and an optical field, generating quantum entanglement between them via gravitational effects. This is achieved by making use of a boson-qubit mapping protocol and a digital gate decomposition that allow us to run the simulations in the quantum computers available in the IBM Quantum platform. We present the obtained results for the fidelity of the experiment in two different quantum computers, after applying error mitigation and post-selection techniques. The achieved results correspond to fidelities over 90\%, which indicates that we were able to perform a faithful digital quantum simulation of the interaction and therefore of the generation of quantum entanglement by gravitational means in optomechanical systems.


\end{abstract}




\end{abstractbox}
%

\end{frontmatter}

\section{Introduction}

 

In the last years a novel approach to the open debate of quantum gravity has emerged, where the focus has shifted to just prove the quantum nature of gravity, without disclosing the underlying full quantum theory \cite{Bose_2017,Marletto_2017,Bose,biswas2023gravitational}.
The key aspect is that, since two quantum systems cannot be entangled via a classical channel if they were not entangled beforehand, if we are able to show that the gravitational field is able to generate entanglement, we can conclude that it must arise from genuine quantum properties, and therefore the field must be a quantum entity.

While the prospects for an experimental implementation of these proposals are encouraging -as compared to a direct proof of the full quantum theory of gravity- actual experimental results are still missing. In this context, a quantum simulation in a quantum computer \cite{Feynman1982Simulating,1996Sci...273.1073L} could be helpful by informing the ongoing experimental efforts, leveraging the experimental amenability and tunability of modern quantum platforms \cite{arute2019quantum,bharti2022noisy,kim2023evidence}.

Some of us have recently introduced a recipe for the digital quantum simulation of bosonic Hamiltonians \cite{Sab_n_2020}, which combines Trotter \cite{1996Sci...273.1073L} and gate-decomposition techniques with a boson-qubit mapping \cite{Somma_2003,somma2005quantum}  to encode the Hamiltonian into a sequence of single-qubit and two-qubit gates. We have applied this scheme to the high-fidelity digital quantum simulation of paradigmatic quantum-optics interactions, such as beam-splitters or single-mode and two-mode squeezers \cite{PhysRevA.104.052609}.

The procedure carried out in this project is similar to the recent work by one of the authors in \cite{Sab_n_2023}, with the main difference laying in the fact that, in that work, the author reports the digital quantum simulation of the Hamiltonian involved in the generation of quantum entanglement between a pair of quantum harmonic oscillators by gravitational means \cite{Bose}, while here we apply this scheme in order to study the quantum gravitational Hamiltonian that represents the interaction of a matter system and an optical field. The study of this Hamiltonian is related to the field of gravitational optomechanics, which is a term that was introduced in \cite{biswas2023gravitational} as a way to refer to any optomechanical interaction that has the gravitational field as its source.

This interaction gives rise to a bosonic two-mode Hamiltonian which we translate into a multiqubit one and then decompose into one and two-qubit quantum gates in order to launch it to IBM quantum devices. Even though the interaction studied in this work is exactly solvable, it provides a framework towards simulating more complex scenarios that could not be studied with analytic treatments. The techniques presented here are the basic building blocks of interactions including more modes and more excitations per mode, which would give rise to those scenarios.

We consider a simplified scenario allowing a maximum of one excitation per mode --a sensible approximation given the weakness of the interaction-- and leverage error mitigation and postselection techniques to achieve fidelities above 90\%. Due to some technical limitations, which are explained in detail later, we were not able to reproduce the current experimental range of parameters, which purely depends on the particular experimental setup that we have studied, so we consider slightly larger values, which amplify the amount of quantum gravitational entanglement generated. Thus, these preliminary four-qubit experiments can be considered as a first step towards larger and more elaborate quantum simulations, with more modes and more photons per mode, beyond the capabilities of classical computers.


The structure of the paper is the following. In the next section, we introduce the Hamiltonian of gravitational optomechanics and develop its digitization. After that, we show how to realize a full quantum state tomography and compute the fidelity of this state with respect to theoretical predictions. Then, we present experimental results for the fidelity of the digital quantum simulation in IBM quantum computers. We conclude with a summary and discussion of our results.

\section{Digitization of gravitational optomechanics}
\label{sec:2}
We are interested in performing the digital quantum simulation of the interaction presented in \cite{biswas2023gravitational} as the one that rules the interaction betweeen a matter-wave and a photonic field mode. This interaction is governed by the following Hamiltonian:
\begin{equation}
    \hat{V}=-g_0(\hat{b}+\hat{b}^\dagger)\hat{a}^\dagger\hat{a},
\label{eqn:ham}
\end{equation}
where $g_0$ is an optomechanical coupling whose value is related to the parameters of the experiment and $\hat{b}^\dagger$ ($\hat{b}$) and $\hat{a}^\dagger$ ($\hat{a}$) are the creation (annihilation) operators of the mechanical oscillator and photon, respectively. What the authors in \cite{biswas2023gravitational} realised is that (\ref{eqn:ham}) is formally the same as the Hamiltonian interaction that arises in the field of optomechanics when studying a quantum system composed of a cavity field interacting with a movable mirror \cite{opto58,opto59}, which means we can apply existing protocols from quantum optomechanics to our interaction. One of these procedures is usually performed when the cavity is strongly driven to some large coherent amplitude $\alpha=\langle\hat{a}\rangle$ \cite{optomechanics}, and it consists in linearising the interaction 
by writing $\hat{a}\rightarrow\alpha+\delta\hat{a}$, where $\delta \hat{a}$ represents the small quantum fluctuations around this mean value $\alpha$. This approximation is only valid when two conditions are met\cite{clarke}, $|\alpha|\gg 1$ and $\mu\sigma\ll1$, where $\mu\sim g_0t$ and $\sigma\sim\sqrt{\text{Var}(\hat{b}+\hat{b}^\dagger)}$. Both of them are fulfilled in our case, so, by renaming $\delta\hat{a}$ as $\hat{a}$, we get:
\begin{equation}
    \hat{H}=-g(\hat{b}+\hat{b}^\dagger)(\hat{a}+\hat{a}^\dagger),
\label{eqn:hamfinal}
\end{equation}
with $g=g_0\alpha$. 
This is formally equivalent to the well-known cavity optomechanical interaction of quantum optics in the linearized regime \cite{opto58,opto59,optomechanics}, but now the coupling is induced by the interaction with the gravitational field encoded in the coupling $g$.


Thus, in order to perform and analyze the simulation of our Hamiltonian, our goal is to develop a quantum circuit that replicates its action when applied on a certain initial state with no entanglement --such as the ground state--, prove that it generates entanglement between the matter and photonic systems and give an estimate of the accuracy of the simulation by calculating the fidelity of the state with respect to an ideal theoretical prediction.

We will now work upon the Hamiltonian presented in (\ref{eqn:hamfinal}). The time evolution of the initial state will be ($\hbar=c=1$):
\begin{equation}
    \ket{\psi(t)}=e^{-iHt}\ket{\psi_0}=e^{igt(\hat{b}+\hat{b}^\dagger)(\hat{a}+\hat{a}^\dagger)}\ket{\psi_0}.
\end{equation}
The main goal of this work will be the digitalization of the following unitary operator: 
\begin{equation}
    U_\varepsilon=e^{i\varepsilon(\hat{b}+\hat{b}^\dagger)(\hat{a}+\hat{a}^\dagger)},
\label{eqn:unitary}
\end{equation}
where we have defined:
\begin{equation}
    \varepsilon=gt.
\label{eqn:epsdef}
\end{equation}
For this purpose, we will need to use the boson mapping techniques presented in \cite{Somma_2003,somma2005quantum}. It is shown there that it is possible to map $N$ bosonic modes with $N_P$ excitations each to $N(N_P+1)$ qubits. In our case, we are dealing with two modes, corresponding to the matter and photon states in (\ref{eqn:hamfinal}). For the rest of this work, we will consider $N_P=1$ for both bosonic modes. We then will need four qubits in all our quantum circuits, which we will label from 0 to 3. Following  \cite{Somma_2003,somma2005quantum}, we will have that:
\begin{equation}
    \begin{split}
    &\ket{0}_m\leftrightarrow\ket{0_01_1}\\
    &\ket{1}_m\leftrightarrow\ket{1_00_1}\\
    &\ket{0}_p\leftrightarrow\ket{0_21_3}\\
    &\ket{1}_p\leftrightarrow\ket{1_20_3},
\end{split}\label{eqn:statesphys}
\end{equation}
where the subscripts $m$ and $p$ refer to the matter and photonic states, respectively. The mapping of the operators for $N_P=1$ leaves us with:
\begin{equation}
\begin{split}
    &\hat{b}^{\dagger}\rightarrow\sigma_-^0\sigma_+^1\\
    &\hat{b}\rightarrow\sigma_+^0\sigma_-^1\\
    &\hat{a}^{\dagger}\rightarrow\sigma_-^2\sigma_+^3\\
    &\hat{a}\rightarrow\sigma_+^2\sigma_-^3.
\end{split}
\end{equation}
Now, we need to express the $(\hat{b}+\hat{b}^{\dagger})$ and the $(\hat{a}+\hat{a}^{\dagger})$ terms in (\ref{eqn:hamfinal}) with respect to the Pauli matrices $\sigma_x,\sigma_y,\sigma_z$. For the first one, we have:
\begin{equation}
\begin{split}
    &\hat{b}^{\dagger}\rightarrow\sigma_-^0\sigma_+^1=\frac14\left(\sigma_x^0- i\sigma_y^0\right)\left(\sigma_x^1+ i\sigma_y^1\right)=\frac14\left(\sigma_x^0\sigma_x^1+i\sigma_x^0\sigma_y^1-i\sigma_y^0\sigma_x^1+\sigma_y^0\sigma_y^1
\right),\\
    &\hat{b}\rightarrow\sigma_+^0\sigma_-^1=\frac14\left(\sigma_x^0+ i\sigma_y^0\right)\left(\sigma_x^1- i\sigma_y^1\right)=\frac14\left(\sigma_x^0\sigma_x^1-i\sigma_x^0\sigma_y^1+i\sigma_y^0\sigma_x^1+\sigma_y^0\sigma_y^1
\right),
\end{split}
\end{equation}
and therefore:
\begin{equation}
    \hat{b}+\hat{b}^{\dagger}\rightarrow\frac12\left(\sigma_x^0\sigma_x^1
+\sigma_y^0\sigma_y^1\right).
\end{equation}
Similarly:
\begin{equation} \hat{a}+\hat{a}^{\dagger}\rightarrow\frac12\left(\sigma_x^2\sigma_x^3
+\sigma_y^2\sigma_y^3\right),
\end{equation}
and finally we get the expression for the full Hamiltonian:
\begin{equation}
\begin{split}
    &\hat{H}=-g(\hat{b}+\hat{b}^\dagger)(\hat{a}+\hat{a}^{\dagger})\rightarrow-\frac{g}{4}\left(\sigma_x^0\sigma_x^1\sigma_x^2\sigma_x^3+\sigma_x^0\sigma_x^1\sigma_y^2\sigma_y^3+\sigma_y^0\sigma_y^1\sigma_x^2\sigma_x^3+\sigma_y^0\sigma_y^1\sigma_y^2\sigma_y^3\right).\\
\end{split}
\label{eqn:hampauli}
\end{equation}

It can be shown that all of the terms in (\ref{eqn:hampauli}) commute with each other, so we can factorize the unitary without the need of Trotter approximations. For instance, let us consider the first and last terms. Clearly, the operators acting on different qubits commute, so we have:
\begin{equation}
    \left[\sigma_x^0\sigma_x^1\sigma_x^2\sigma_x^3,\sigma_y^0\sigma_y^1\sigma_y^2\sigma_y^3\right]=\left[\sigma_x^0\sigma_x^1,\sigma_y^0\sigma_y^1\right]\sigma_x^2\sigma_x^3\sigma_y^2\sigma_y^3+\sigma_x^0\sigma_x^1\sigma_y^0\sigma_y^1\left[
\sigma_x^2\sigma_x^3,
\sigma_y^2\sigma_y^3\right].
\label{eqn:comgrande}
\end{equation}
In a similar way:
\begin{equation}
\begin{split}
&\left[\sigma_x^0\sigma_x^1,\sigma_y^0\sigma_y^1\right]=\left[\sigma_x^0,\sigma_y^0\right]\sigma_y^1
\sigma_x^1+\sigma_x^0
\sigma_y^0\left[\sigma_x^1,\sigma_y^1\right]\\
&\left[\sigma_x^2\sigma_x^3,\sigma_y^2\sigma_y^3\right]=\left[\sigma_x^2,\sigma_y^2\right]\sigma_y^3
\sigma_x^3+\sigma_x^2
\sigma_y^2\left[\sigma_x^3,\sigma_y^3\right].
\end{split}    
\end{equation}
Recalling the commutation relationships of Pauli matrices, we get:
\begin{equation}
    \left[\sigma_x^{(k)}\sigma_x^{(l)},\sigma_y^{(k)}\sigma_y^{(l)}\right]=\left[\sigma_x^{(k)},\sigma_y^{(k)}\right]\sigma_y^{(l)}
\sigma_x^{(l)}+\sigma_x^{(k)}
\sigma_y^{(k)}\left[\sigma_x^{(l)},\sigma_y^{(l)}\right]=2\sigma_z^{(k)}\sigma_z^{(l)}-2\sigma_z^{(k)}\sigma_z^{(l)}=0,
\end{equation}
and thus the two terms in (\ref{eqn:comgrande}) vanish and so does the full commutator. An equivalent procedure can be carried out for the rest of the terms. Therefore we can write:
\begin{equation}
    U_\varepsilon=\prod_{i=1}^4U_\varepsilon^{(i)}=\prod_{i=1}^4e^{-i\hat{H}_it},
\end{equation}
where the $U_\varepsilon^{(i)}$ are obtained as the exponential of each of the terms in (\ref{eqn:hampauli}). For instance:
\begin{equation}
    U_\varepsilon^{(1)}=e^{i\tilde{\varepsilon}\sigma_x^0\sigma_x^1\sigma_x^2\sigma_x^3},
\end{equation}
with $\tilde{\varepsilon}=\varepsilon/4$. In principle, these are already unitary gates that we could implement in a quantum circuit to simulate our Hamiltonian. However, these unitary gates are not available in the IBM quantum devices, which means that we have to carry out a gate decomposition to express each of them in terms of gates that are actually available in the IBM software \cite{Sab_n_2023,PhysRevA.104.052609,Sab_n_2020}.

Let $U$ be some unitary operation such that, for a Hamiltonian $H$, it satisfies:
\begin{equation}
    H=U^\dagger H_0U,
\label{eqn:dec1}
\end{equation}
where $H_0$ is some convenient single-qubit operation. Then, we can always write:
\begin{equation}
    e^{-iH\tilde{\varepsilon}}=U^\dagger e^{-iH_0\tilde{\varepsilon}}U.
\end{equation}
Our goal is now to carry out this procedure for each piece of our Hamiltonian. For example, for $U_\varepsilon^{\left(1\right)}$, by defining $H_0=-\sigma_z^0$, we have that, using (\ref{eqn:dec1}):

\begin{equation}
\begin{split}
    &e^{-i\frac{\pi}{4}\sigma_x^0}(-\sigma_z^0)e^{i\frac{\pi}{4}\sigma_x^0}=\sigma_y^0\\
    &\\
    &e^{-i\frac{\pi}{4}\sigma_z^0\sigma_x^1}\sigma_y^0 e^{i\frac{\pi}{4}\sigma_z^0\sigma_x^1}=-\sigma_x^0\sigma_x^1\\
    &\\
    &e^{-i\frac{\pi}{4}\sigma_z^0\sigma_x^2}(-\sigma_x^0\sigma_x^1) e^{i\frac{\pi}{4}\sigma_z^0\sigma_x^2}=-\sigma_y^0\sigma_x^1\sigma_x^2\\
    &\\
    &e^{-i\frac{\pi}{4}\sigma_z^0\sigma_x^3}(-\sigma_y^0\sigma_x^1\sigma_x^2) e^{i\frac{\pi}{4}\sigma_z^0\sigma_x^3}=\sigma_x^0\sigma_x^1\sigma_x^2\sigma_x^3.
\end{split}
\end{equation}
Thus, we find that  $U_\varepsilon^{\left(1\right)}$ can be rewritten as:
\begin{equation}
    U_\varepsilon^{\left(1\right)}=U^\dagger e^{i\tilde{\varepsilon}\sigma_z^0}U,
\end{equation}
with 
\begin{equation}
    U=e^{i\frac{\pi}{4}\sigma_x^0}
e^{i\frac{\pi}{4}\sigma_z^0\sigma_x^1}e^{i\frac{\pi}{4}\sigma_z^0\sigma_x^2}e^{i\frac{\pi}{4}\sigma_z^0\sigma_x^3}.
\label{eqn:opU}
\end{equation}
The decompositions for the rest of the $U_\varepsilon^{\left(i\right)}$ can be found by adding the necessary $e^{i\pi/4\sigma_z^{(i)}}$ to the operator in (\ref{eqn:opU}), in order to turn the corresponding $\sigma_x$ into $\sigma_y$ in each case. 
As stated earlier, the purpose of this gate decomposition technique was to be able to express our Hamiltonian operator in terms of gates that can be implemented in the IBM quantum computers. Taking a look
at the decomposition of each of our terms (see Table 1), it can be noticed that there are only three different types of operators that were needed throughout the whole process: The ones with $\sigma_x^i$, $\sigma_z^j$ or $\sigma_x^i\sigma_z^j$ in the exponent. All three of these operators can be converted into gates available in the IBM architecture \cite{Sab_n_2023,PhysRevA.104.052609}.  For the first two, we can use the single-qubit rotations $R_x(\theta)$ and $U_1(\lambda)$:
\begin{equation}
    R_x(\theta)=e^{-i\frac{\theta}{2}\sigma_x}
\end{equation}
\begin{equation}
    U_1(\lambda)=\begin{pmatrix}
    1 & 0 \\
    0 & e^{i\lambda} 
    \end{pmatrix}=e^{i\frac{\lambda}{2}}e^{-i\frac{\lambda}{2}\sigma_z},
\end{equation}
where the $U_1$ gate takes the particular form of the $S$ gate for $\lambda=\pi/2$:
\begin{equation}
    S=\begin{pmatrix}
    1 & 0 \\
    0 & i 
    \end{pmatrix}=e^{i\frac{\pi}{4}}e^{-i\frac{\pi}{4}\sigma_z}.
\end{equation}
For the last type of gates, the conversion is given by:
\begin{equation}
    e^{i\frac{\pi}{4}\sigma_z^i\sigma_x^j}=e^{i\frac{\pi}{4}\sigma_x^j}e^{i\frac{\pi}{4}\sigma_z^i}e^{-i\frac{\pi}{4}}\text{CNOT}^{i-j},
\end{equation}
where the CNOT gate is available in IBM for some pairs of qubits, depending on the particular device connectivity.

We can now put all this together to express each of the terms of the unitary with a combination of these three types of gates. For $U_\varepsilon^{(1)}$, we will have that, up to an irrelevant global phase and after some cancellations and simplifications:
\begin{equation}
\begin{split}
    &U_\varepsilon^{\left(1\right)}=\text{CNOT}^{0-3}S^{0}\text{CNOT}^{0-2}S^{0}\text{CNOT}^{0-1}S^{0}R_x^0\left(\frac{\pi}{2}\right)U_1^0\left(\frac{\varepsilon}{2}\right)R_x^0\left(-\frac{\pi}{2}\right)S^{\dagger 0}\text{CNOT}^{0-1}\\
    &S^{\dagger 0}\text{CNOT}^{0-2}S^{\dagger 0}\text{CNOT}^{0-3}.
\end{split}
\end{equation}
For the sake of completeness, we leave the rest of the digital decompositions for the $U_\varepsilon^{\left(i\right)}$ in Appendix A.

It is also worth mentioning that when we send a quantum circuit to an IBM quantum computer, the circuit where the measurements are actually performed is usually not exactly the same as the one we designed, since the circuit is first transpiled, that is, optimized to the particular quantum architecture used. 

The gate set of the computers that our computations are being carried out in can be checked accessing their configuration. All of the accessible quantum computers possess the same gate set, formed by the X, CNOT, SX (also referred to as $\sqrt{X}$, and equivalent to a $R_x(\pi/2)$ gate) and $R_z$ gates, on top of the identity and the reset gates. Therefore, every single gate we place in our circuit will be transformed to a combination of some of these gates. 

The transpilation of a circuit might increase or decrease the number of gates that are needed to fulfill the action of the circuit with respect to the original one. We are in general interested in having the circuit simplified after the transpilation, in order to reduce the number of gates and therefore the global error. However, connectivity issues might translate into additional gates.

\begin{figure*}[h!]
\centering
\includegraphics[width=0.9\textwidth]{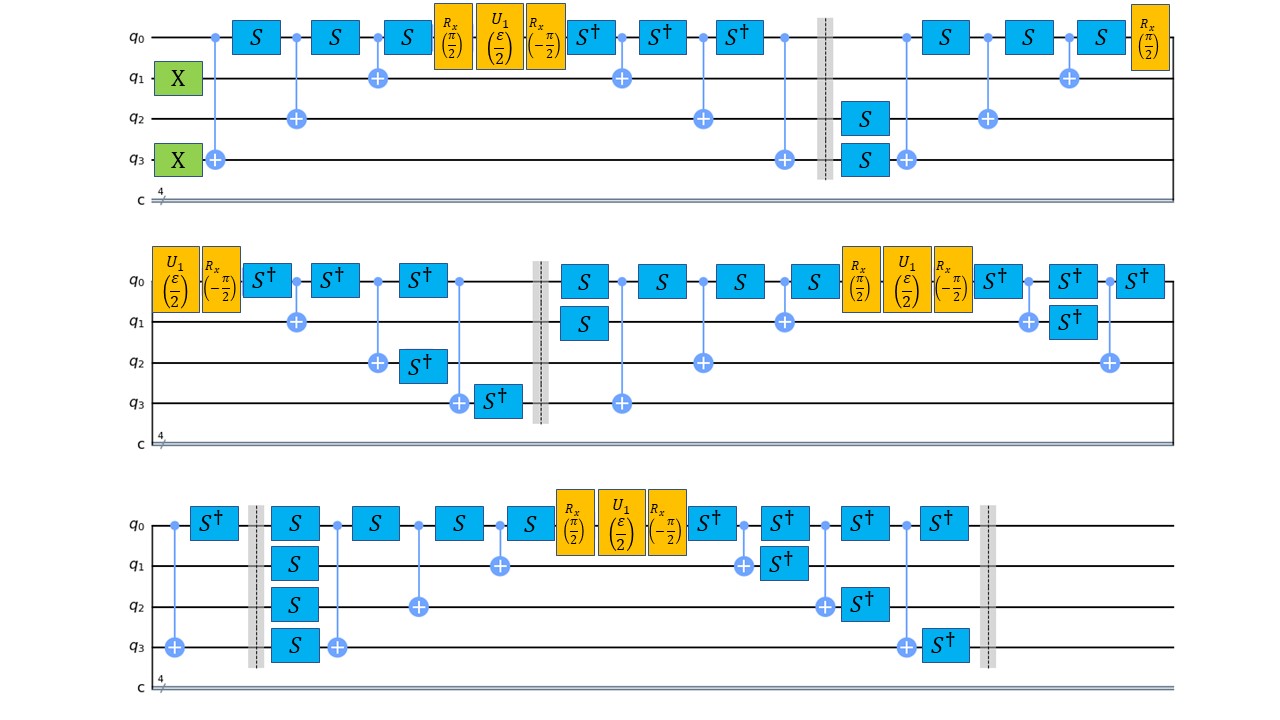}
\caption{Quantum circuit for the simulation of $U_\varepsilon$.}
\label{fig:circuit}
\end{figure*}

\begin{figure*}[h!]
\centering
\includegraphics[width=0.9\textwidth]{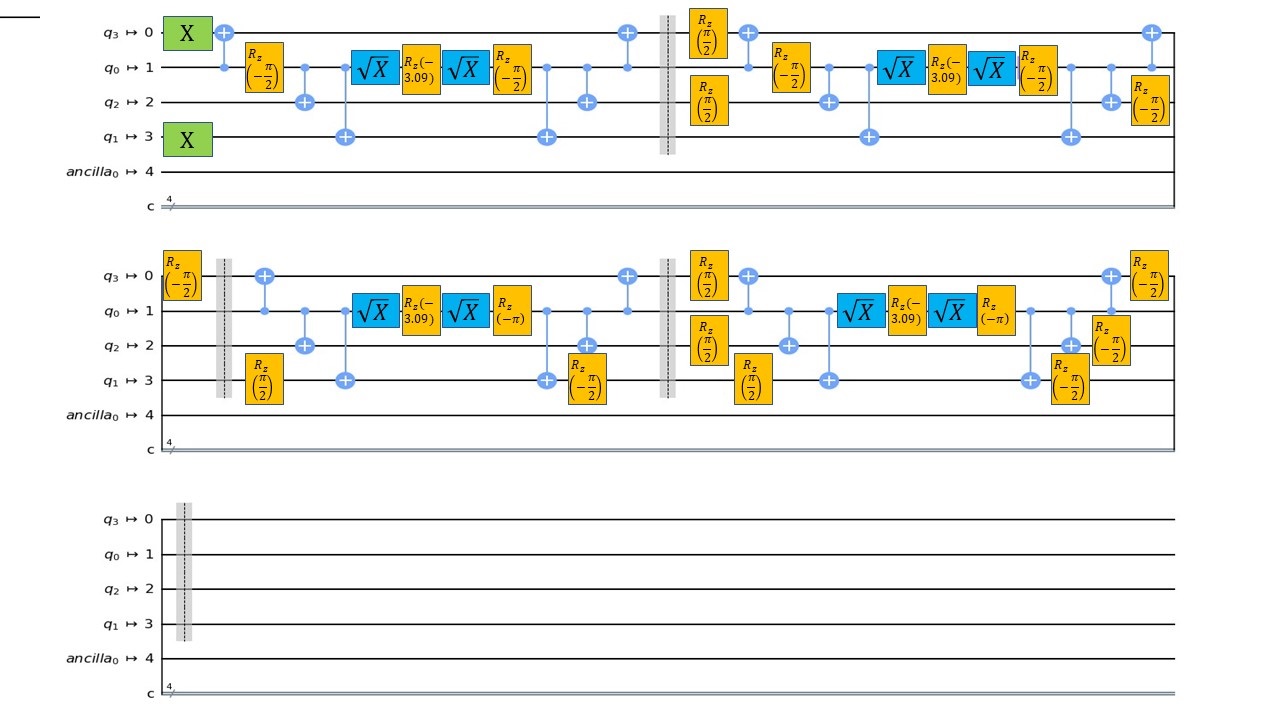}
\caption{Transpiled quantum circuit in IBM Belem for $\varepsilon=0.1$.}
\label{fig:transpiled_b}
\end{figure*}

\begin{figure}[h!]
\centering
\includegraphics[width=0.9\textwidth]{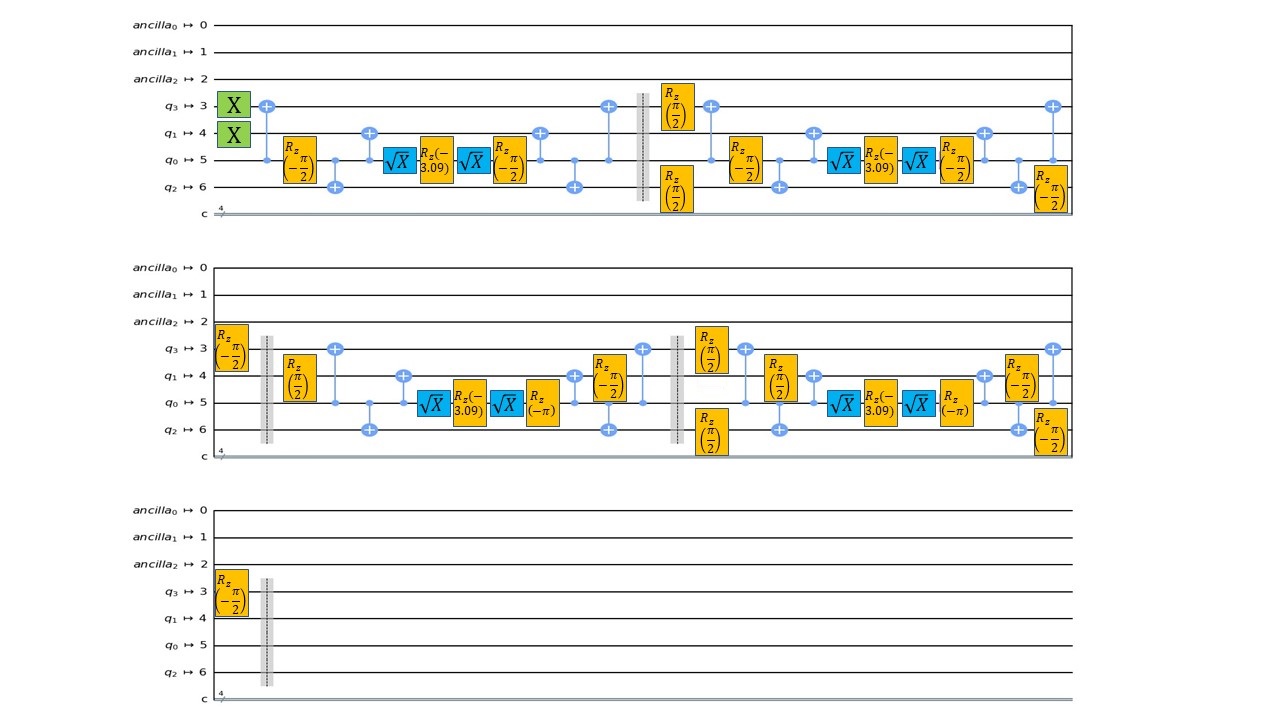}
\caption{Transpiled quantum circuit in IBM Nairobi for $\varepsilon=0.1$.}
\label{fig:transpiled_n}
\end{figure}

In Figures \ref{fig:circuit}, \ref{fig:transpiled_b} and \ref{fig:transpiled_n}, we show the full quantum circuit that is implemented to simulate the action of our time evolution operator and the transpiled version of said circuit in the quantum computers IBM Belem and IBM Nairobi. The transpiled circuits look different due to the different number of qubits and connectivity. It is also worth remarking that the value of $\varepsilon$ does not affect the structure of the original quantum circuit or the transpiled versions, but only results in changes in the parameters in some of the quantum gates.

\captionsetup[figure]{font=footnotesize,labelfont=footnotesize}
\captionsetup[figure]{font=footnotesize,labelfont=footnotesize}

Note that the initial two X-gates in Figure \ref{fig:circuit} are introduced in order to prepare the system into the initial state which we worked with, which is the ground state:
\begin{equation}
    \ket{0}_m\ket{0}_p\rightarrow\ket{0_01_10_21_3}.
\end{equation}
By observing these three images, we can see that the transpilation process helps slightly simplify the circuit by decreasing the total number of single-qubit gates required to simulate the action of $U_\varepsilon$ without increasing the number of two-qubit gates. For both quantum computers, the circuit consists of 32 single-qubit gates and 24 CNOT gates after the transpilation.

\section{Fidelity}
\label{subsec:fid}
In order to determine the quality of our digitalization, we calculate the fidelity as \cite{Nielsen,Sab_n_2023,PhysRevA.104.052609}:
\begin{equation}
    F\left(\ket{\psi},\rho\right)=\bra{\psi}\rho\ket{\psi},
\label{eqn:fidelity}
\end{equation}
where $\ket{\psi}$ is the state at which the system will be after the action of the full quantum circuit (that is, after the action of $U_\varepsilon$), while $\rho$ will be the density matrix that represents the state actually obtained in the experiment. We will now analyze the procedure that is required to obtain an expression for the fidelity as a function of $\varepsilon$.

Considering again $U_\varepsilon$, if we restrict ourselves to perturbative values of $\varepsilon$, we can perform a Taylor expansion of $U_\varepsilon$ up to second order in $\varepsilon$ and provide an estimate to the state of the system after a certain time, by applying this expression for $U_\varepsilon$ to our initial state $\ket{\psi_0}$. We arrive to:

\begin{equation}
    \ket{\psi}=\left(1-\frac{\varepsilon^2}{2}\right)\ket{0}_m\ket{0}_p+i\varepsilon
\ket{1}_m
\ket{1}_p, 
\label{eqn:estado}
\end{equation}
which is, as we can see, an entangled state involving the ground and first excited states of the matter system and the optical field. In particular its concurrence, defined for pure states as \cite{Rungta_2001,Hill_1997}:
\begin{equation}
    C\equiv\sqrt{2\left(1-\text{tr}\left(\sigma^2_m\right)\right)},
\end{equation}
where $\sigma_m$ is the reduced density matrix of the matter oscillator subsystem, can be quiqkly found to be, up to second order:
\begin{equation}
    C=2\varepsilon=2gt. \label{eqn:conc}
\end{equation}
Let us now consider the experimental density operator. We perform a full state tomography of $\rho$. 
For a two qubit state, we have \cite{Nielsen}:
\begin{equation}
    \rho=\frac14\sum_{\vec{v}}\text{tr}\left(\sigma_{v_1}\otimes\sigma_{v_2}\rho\right)\sigma_{v_1}\otimes\sigma_{v_2},
\end{equation}
where the sum is over vectors $\vec{v}=(v_1,...,v_n)$ with entries $v_i$ chosen from the set $0,1,2,3$. Plugging this into (\ref{eqn:fidelity}), we have:
\begin{equation}
    F\left(\ket{\psi},\rho\right)=\frac14\sum_{i,j}\text{tr}\left(\sigma_{i,m}\otimes\sigma_{j,p}\rho\right)\bra{\psi}\sigma_{i,m}\otimes\sigma_{j,p}\ket{\psi},\label{eqn:fidexp}
\end{equation}
where $i,j=0,1,2,3$ and $\sigma_0$ is defined as the identity matrix. Using eq. (\ref{eqn:estado}) we obtain that the nonzero terms in eq. (\ref{eqn:fidexp}) are:
\begin{equation}
    \begin{split}
        &\bra{\psi}\mathbb{1}_{m}\otimes\mathbb{1}_{p}\ket{\psi}=1\\
        &\bra{\psi}\mathbb{1}_{m}\otimes\sigma_{z,p}\ket{\psi}=1-2\varepsilon^2\\
        &\bra{\psi}\sigma_{z,m}\otimes\mathbb{1}_{p}\ket{\psi}=1-2\varepsilon^2\\
        &\bra{\psi}\sigma_{z,m}\otimes\sigma_{z,p}\ket{\psi}=1\\
        &\bra{\psi}\sigma_{x,m}\otimes\sigma_{y,p}\ket{\psi}=2\varepsilon\\
        &\bra{\psi}\sigma_{y,m}\otimes\sigma_{x,p}\ket{\psi}=2\varepsilon,
    \end{split}
\end{equation}
which leads us to:
\begin{equation}
    \begin{split}
    &F\left(\ket{\psi},\rho\right)=\frac14\Big(1+\text{tr}\left(\sigma_{z,m}\otimes\sigma_{z,p}\rho\right)+2\varepsilon\left(\text{tr}\left(\sigma_{y,m}\otimes\sigma_{x,p}\rho\right)+\text{tr}\left(\sigma_{x,m}\otimes\sigma_{y,p}\rho\right)\right)\\
    &+\left(1-2\varepsilon^2\right)\left(\text{tr}\left(\mathbb{1}_{m}\otimes\sigma_{z,p}\rho\right)+\text{tr}\left(\sigma_{z,m}\otimes\mathbb{1}_{p}\rho\right)\right)\Big).
    \end{split}
\label{eqn:fidelityfinal}
\end{equation}
The traces in eq. (\ref{eqn:fidelityfinal}) are expected values of observables of the form $\sigma_{i,m}\otimes\sigma_{j,p}$ over $\rho$, and we can express them as \cite{PhysRevA.104.052609}:
\begin{equation}
    \begin{split}   
    &\text{tr}\left(\sigma_{z,m}\otimes\sigma_{z,p}\rho\right)=p_{ZZ}(\ket{0}_m\ket{0}_p)+p_{ZZ}(\ket{1}_m\ket{1}_p)-p_{ZZ}(\ket{0}_m\ket{1}_p)-p_{ZZ}(\ket{1}_m\ket{0}_p)\\
    &\text{tr}\left(\sigma_{y,m}\otimes\sigma_{x,p}\rho\right)=p_{YX}(\ket{0}_m\ket{0}_p)+p_{YX}(\ket{1}_m\ket{1}_p)-p_{YX}(\ket{0}_m\ket{1}_p)-p_{YX}(\ket{1}_m\ket{0}_p)\\
    &\text{tr}\left(\sigma_{x,m}\otimes\sigma_{y,p}\rho\right)=p_{XY}(\ket{0}_m\ket{0}_p)+p_{XY}(\ket{1}_m\ket{1}_p)-p_{XY}(\ket{0}_m\ket{1}_p)-p_{XY}(\ket{1}_m\ket{0}_p)\\
    &\text{tr}\left(\mathbb{1}_{m}\otimes\sigma_{z,p}\rho\right)=p_{Z}(\ket{0}_p)-p_{Z}(\ket{1}_p)\\
    &\text{tr}\left(\sigma_{z,m}\otimes\mathbb{1}_{p}\rho\right)=p_{Z}(\ket{0}_m)-p_{Z}(\ket{1}_m),\\
    \end{split}
\label{eqn:traces}
\end{equation}
where these quantities are the probabilities of each mode being in state $\ket{0}$ or $\ket{1}$ of the corresponding basis in each case. For instance, $p_{XY}(\ket{0}_m\ket{0}_p)$ represents the probability of the matter and photon systems of being in the $\ket{0}$ state of bases $X$ and $Y$ respectively. Nonetheless, the measuring operator available in the IBM system is equivalent to measuring in the $Z$ basis and there is no way to perform a measure directly in the other two. This can be solved by implementing quantum gates in the relevant qubits before the measurement, flipping them to $X$ or $Y$ basis. In particular, we apply a Hadamard gate before measurement in the case of $X$-measurements, and $S^\dagger$ and Hadamard  for $Y$-measurements.
~~~~~~~~~~~~~~~~~~~~~~~~~~~~~~~~~~~~~~~~

With this, we are already able to perform all the necessary measurents for the computation of the fidelity of our experiment. 
Moreover, in order to improve the fidelity, we apply error mitigation in the readout measurements and postselect the results to the subspace of the 4-qubit Hilbert space with physical meaning in our simulation, namely the one spanned by the four states in (\ref{eqn:statesphys}). 

\section{Results}

We have now developed all the necessary tools to implement the quantum simulation of the Hamiltonian in (\ref{eqn:hamfinal}), but we still must consider an appropriate range of values of $\varepsilon$.

\begin{figure}[h!]
    \centering
    \begin{subfigure}{0.47\textwidth}
        \centering
        \includegraphics[width=\textwidth]{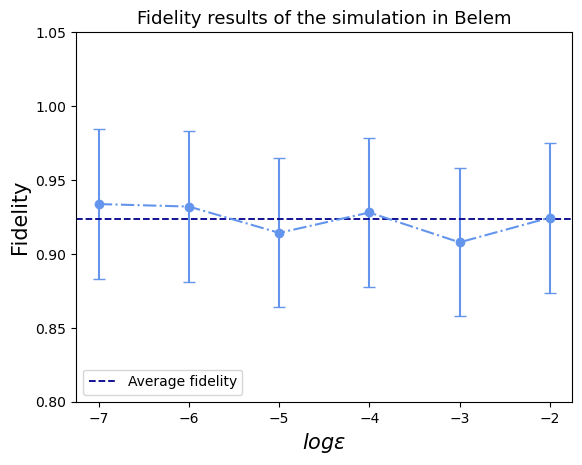}
    \end{subfigure}
    \hfill
    \begin{subfigure}{0.47\textwidth}
        \centering
        \includegraphics[width=\textwidth]{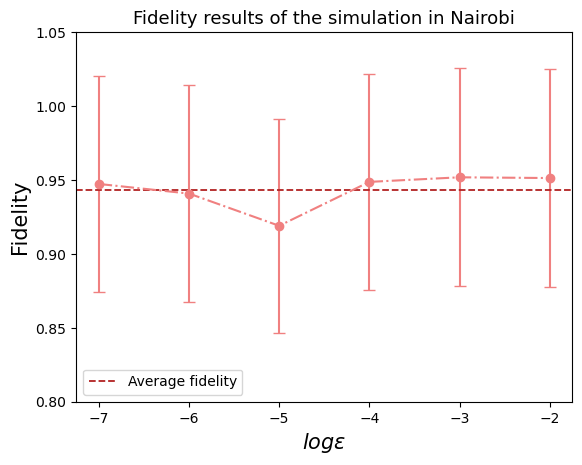}
   \end{subfigure}
    \caption{Fidelity results of the digital quantum simulation procedure in the IBM Belem and IBM Nairobi computers.}
    \label{fig:BN}
\end{figure}

From (\ref{eqn:epsdef}), we recall that the value of $\varepsilon$ is determined by that of the gravitational optomechanical coupling $g$ in the Hamiltonian and the experimental time for which we are assuming  we would let the system evolve. In \cite{biswas2023gravitational}, it is mentioned that, with the currently available experimental setups, a value of $2\pi\cdot 10^{-11}$ for $g$ could be reached and adequate values of $t$ would be of the order of seconds, giving rise to a value of $\varepsilon$ around $10^{-10}$. However, IBM's Qiskit approximates to 0 any number smaller than $10^{-9}$.
Notice, however, that the amount of entanglement generated by the quantum gravitational field is proportional to $\varepsilon$, according to eq. (\ref{eqn:conc}). Thus, it would be of interest to simulate the system for values of $\varepsilon$ that magnify the generated entanglement while lying beyond current experimental reach --but remaining in the perturbative regime.
In Figure \ref{fig:BN}, we show the results for the fidelity of the state after the action of the circuit for values of $\varepsilon$ that range from $10^{-7}$ to $10^{-2}$. The corresponding measurements were carried out in two different IBM quantum computers, namely Belem and Nairobi, with 5 and 7 qubits and 16 and 32 quantum volume \cite{Cross_2019} respectively. The average error rates of the IBM Belem (Nairobi) quantum computer are of around $2.76~(3.28)\times 10^{-4}$ for the single-qubit gates, $8.75~(14.92) \times 10^{-3}$   for the CNOT gates and $2.11~(3.06)\times 10^{-2}$  for the readout --however, recall that we have made use of an error mitigation technique for our measurements. It is important to remark that these parameters change on a daily basis. The experiments in Belem were carried out on June 10, 2023, while the ones in Nairobi were done on June 18, 2023. The error bars in the graphs are related to the readout error and the way they are computed is detailed in Appendix B. 

\captionsetup[figure]{font=footnotesize,labelfont=footnotesize}

As we can see, we have achieved values for the fidelity between 0.9 and 0.95 in both cases, after applying error mitigation and postselection. The results are improved to a small extent when the measurements are done in the IBM Nairobi computer, raising the average fidelity from around 92\% to 94\%, due to its larger quantum volume.

\section{Conclusions}

In this work, we carry out the digital quantum simulation of a Hamiltonian governing the interaction between  a mechanical oscillator and an optical field which interact through a quantum gravitational field generating entanglement between them. Therefore, we simulate an experiment for the generation of gravitational entanglement to prove the quantum nature of gravity.

This is achieved by implementing the corresponding boson-qubit mapping  and performing a suitable gate decomposition in order to write our time evolution operator $U_\varepsilon$ in terms of gates that can be used in the IBM Quantum platform computers. Exploiting post-selection and error mitigation techniques, we achieve high-fidelity --above 90\%-- in two different devices, and observe an improvement of the fidelity in the device with larger quantum volume. Given the rate at which IBM is increasing the quantum volume, it could soon be possible  to increase the number of qubits while keeping high fidelity. This would allow us to relax the restriction to one maximum excitation per mode. While this should be a very good approximation given the weakness of the coupling, increasing the allowed excitations and thus the number of qubits will enable to reach the post-classical regime, going beyond the capabilities of classical computers. As mentioned in the introduction, this is the reason why we found that it would be convenient to carry out this simulation protocol in a quantum computer: While the interaction that was presented here can be studied analytically, it would be computationally harder to do so with more complex versions of the process, either with more modes or excitations per mode. For instance, with the currently available 127-qubit computers, we could consider more than 50 excitations per mode, which would also allows us to consider larger coupling values and therefore simulate the gravitational entanglement generated by large gravitational fields. While the quantum volume of these devices have already raised to 128, it is likely that it would not be enough to compensate for the dramatic improvement in the number of quantum gates that would be required to simulate such a large number of excitations. However, novel and more sophisticate error mitigation techniques are currently available \cite{gupta2023probabilistic}, enabling error mitigation in the full circuit and not only in the readout, which would presumably improve the fidelity of those large post-classical experiments. Meanwhile, our preliminary four-qubit results represent a first step along that direction, since larger simulations would essentially require the same techniques described here.

\section*{Appendices}
\subsection*{Appendix A: Gate decomposition}
\label{subsec:appA}
The decompositions of the $U_\varepsilon^{\left(i\right)}$ in terms of gates that are available in the IBM software are:
\begin{equation}
\begin{split}
&U_\varepsilon^{\left(1\right)}=\text{CNOT}^{0-3}S^{0}\text{CNOT}^{0-2}S^{0}\text{CNOT}^{0-1}S^{0}R_x^0\left(\frac{\pi}{2}\right)U_1^0\left(\frac{\varepsilon}{2}\right)R_x^0\left(-\frac{\pi}{2}\right)S^{\dagger 0}\text{CNOT}^{0-1}\\
&S^{\dagger 0}\text{CNOT}^{0-2}S^{\dagger 0}\text{CNOT}^{0-3}.\\
&\\
&U_\varepsilon^{\left(2\right)}=S^{3}S^{2}\text{CNOT}^{0-3}S^{0}\text{CNOT}^{0-2}S^{0}\text{CNOT}^{0-1}S^{0}R_x^0\left(\frac{\pi}{2}\right)U_1^0\left(\frac{\varepsilon}{2}\right)R_x^0\left(-\frac{\pi}{2}\right)S^{\dagger 0}\\
&\text{CNOT}^{0-1}S^{\dagger 0}\text{CNOT}^{0-2}S^{\dagger 0}\text{CNOT}^{0-3}S^{\dagger 2}S^{\dagger 3}.\\
&\\
&U_\varepsilon^{\left(3\right)}=S^{1}S^{0}\text{CNOT}^{0-3}S^{0}\text{CNOT}^{0-2}S^{0}\text{CNOT}^{0-1}S^{0}R_x^0\left(\frac{\pi}{2}\right)U_1^0\left(\frac{\varepsilon}{2}\right)R_x^0\left(-\frac{\pi}{2}\right)S^{\dagger 0}\\
&\text{CNOT}^{0-1}S^{\dagger 0}\text{CNOT}^{0-2}S^{\dagger 0}\text{CNOT}^{0-3}S^{\dagger 0}S^{\dagger 1}.\\
&\\
&U_\varepsilon^{\left(4\right)}=S^{3}S^{2}S^{1}S^{0}\text{CNOT}^{0-3}S^{0}\text{CNOT}^{0-2}S^{0}\text{CNOT}^{0-1}S^{0}R_x^0\left(\frac{\pi}{2}\right)U_1^0\left(\frac{\varepsilon}{2}\right)R_x^0\left(-\frac{\pi}{2}\right)\\
&S^{\dagger 0}\text{CNOT}^{0-1}S^{\dagger 0}\text{CNOT}^{0-2}S^{\dagger 0}\text{CNOT}^{0-3}S^{\dagger 0}S^{\dagger 1}S^{\dagger 2}S^{\dagger 3}.
\end{split}
\end{equation}

\subsection*{Appendix B: Fidelity error}
\label{subsec:appC}

In order to estimate the error of the fidelity for each value of $\varepsilon$, we recall the expression in (\ref{eqn:fidelityfinal}), in which we express the fidelity in terms of probabilities that are determined experimentally by performing measurements on the circuit in different bases. These measurements have an associated error, which we can use to estimate the error in the fidelity through standard error propagation:
\begin{equation}
    \Delta F=\left(\sum_i\left(\dfrac{\partial F}{\partial (\text{tr}_{i}\rho)}\Delta (\text{tr}_{i}\rho)\right)^2\right)^{1/2},
\end{equation}
As a good approximation, we can consider that the readout error is the same for all qubits -the averager readout error provided by IBM- and doesn't depend on whether the measurement is meant to result in 0 or 1. Let us refer to this magnitude as $\lambda$. Since these erorr rates are small, the rate of error in the process of measuring $n$ qubits can be taken, in the perturbative regime, as $\delta=n\cdot\lambda$. Therefore, every time that we perform N measures on the circuit to calculate each of these probabilities, the number of ``counts" that carry along a mistake in the readout will be $\Delta=\delta\cdot N$.

Let us now consider the $\text{tr}_{ZZ}\rho$ term as an example. It was calculated as:
\begin{equation}
    \text{tr}_{ZZ}\rho=\frac{N+N'-N''-N'''}{N+N'+N''+N'''}\equiv\frac{a}{b},
\end{equation}
where the $N$ represent the counts associated to each of the probabilities that are calculated to estimate the value of the trace. Through linear error propagation, we can see that the error associated to both the numerator and the denominator will be $\Delta_{tot}\equiv\Delta a=\Delta b=\Delta+\Delta'+\Delta''+\Delta'''$. The total error associated to this term will be:
\begin{equation}
    \Delta(\text{tr}_{ZZ}\rho)=\bigg|\dfrac{\partial \text{tr}_{ZZ}\rho}{\partial a}\bigg|\Delta a+\bigg|\dfrac{\partial \text{tr}_{ZZ}\rho}{\partial b}\bigg|\Delta b=\frac{a}{b}\Delta_{tot}\left(\frac{1}{a}+\frac{1}{b}\right)=\frac{a}{b}\delta N_{tot}\left(\frac{1}{a}+\frac{1}{b}\right),
\end{equation}
with $N_{tot}=N+N'+N''+N'''=b$. Thus:
\begin{equation}
    \Delta(\text{tr}_{ZZ}\rho)=\delta+\delta\frac{a}{b}=\delta\left(1+\text{tr}_{ZZ}\rho\right)=n\lambda(1+\text{tr}_{ZZ}\rho).
\end{equation}
Finally, we can write an expression for the uncertainty in the fidelity, which is what is displayed as the error bars in the figures:
\begin{equation}
    \Delta F=\frac{1}{4}\sqrt{\Delta(\text{tr}_{ZZ}\rho)^2 + 4\varepsilon^2 \left(\Delta(\text{tr}_{XY}\rho)^2 + \Delta(\text{tr}_{YX}\rho)^2\right) + (1-4\varepsilon^2)\left(\Delta(\text{tr}_{ZI}\rho)^2 + \Delta(\text{tr}_{IZ}\rho)^2\right)}.
\end{equation}
\newpage
\begin{backmatter}

\section*{Declarations}

\section*{Acknowledgements}

\section*{Abbreviations}
P. G. C. R.: Pablo Guillermo Carmona Rufo
A. M.: Anupam Mazumdar
S. B.: Sougato Bose
C. S.: Carlos Sabín

\section*{Funding}
P. G. C. R. acknowledges:  Grant PRE2022-102488 funded by MCIN/AEI/10.13039/501100011033 and FSE+. Project code: PID2021-127726NB-I00.

C.S. acknowledges financial support through the Ramón y Cajal Programme (RYC2019-028014-I).

%
\section*{Availability of data and materials}
Not applicable.
\section*{Ethics approval and consent to participate}
Not applicable.
\section*{Competing interests}
The authors declare that they have no competing interests.

\section*{Consent for publication}
Not applicable

\section*{Authors' contributions}
AM and SB suggested the project. PGCR made the computations, programmed and launched the experiments, analysed the data and wrote the first version of the manuscript. CS supervised the project. All the authors contributed to the final writing of the manuscript.


\newcommand{\BMCxmlcomment}[1]{}

\BMCxmlcomment{

<refgrp>

<bibl id="B1">
  <title><p>Spin Entanglement Witness for Quantum Gravity</p></title>
  <aug>
    <au><snm>Bose</snm><fnm>S</fnm></au>
    <au><snm>Mazumdar</snm><fnm>A</fnm></au>
    <au><snm>Morley</snm><fnm>GW</fnm></au>
    <au><snm>Ulbricht</snm><fnm>H</fnm></au>
    <au><snm>Toro{\v{s} }</snm><fnm>M</fnm></au>
    <au><snm>Paternostro</snm><fnm>M</fnm></au>
    <au><snm>Geraci</snm><fnm>AA</fnm></au>
    <au><snm>Barker</snm><fnm>PF</fnm></au>
    <au><snm>Kim</snm><fnm>M</fnm></au>
    <au><snm>Milburn</snm><fnm>G</fnm></au>
  </aug>
  <source>Physical Review Letters</source>
  <publisher>American Physical Society ({APS})</publisher>
  <pubdate>2017</pubdate>
  <volume>119</volume>
  <issue>24</issue>
  <url>https://doi.org/10.11032Fphysrevlett.119.240401</url>
</bibl>

<bibl id="B2">
  <title><p>Gravitationally Induced Entanglement between Two Massive Particles is Sufficient Evidence of Quantum Effects in Gravity</p></title>
  <aug>
    <au><snm>Marletto</snm><fnm>C.</fnm></au>
    <au><snm>Vedral</snm><fnm>V.</fnm></au>
  </aug>
  <source>Physical Review Letters</source>
  <publisher>American Physical Society (APS)</publisher>
  <pubdate>2017</pubdate>
  <volume>119</volume>
  <issue>24</issue>
  <url>http://dx.doi.org/10.1103/PhysRevLett.119.240402</url>
</bibl>

<bibl id="B3">
  <title><p>Mechanism for the quantum natured gravitons to entangle masses</p></title>
  <aug>
    <au><snm>Bose</snm><fnm>S</fnm></au>
    <au><snm>Mazumdar</snm><fnm>A</fnm></au>
    <au><snm>Schut</snm><fnm>M</fnm></au>
    <au><snm>\v{s}\fi{}</snm><fnm>M</fnm></au>
  </aug>
  <source>Phys. Rev. D</source>
  <publisher>American Physical Society</publisher>
  <pubdate>2022</pubdate>
  <volume>105</volume>
  <fpage>106028</fpage>
  <url>https://link.aps.org/doi/10.1103/PhysRevD.105.106028</url>
</bibl>

<bibl id="B4">
  <title><p>Gravitational Optomechanics: Photon-Matter Entanglement via Graviton Exchange</p></title>
  <aug>
    <au><snm>Biswas</snm><fnm>D</fnm></au>
    <au><snm>Bose</snm><fnm>S</fnm></au>
    <au><snm>Mazumdar</snm><fnm>A</fnm></au>
    <au><snm>Toroš</snm><fnm>M</fnm></au>
  </aug>
  <pubdate>2023</pubdate>
</bibl>

<bibl id="B5">
  <title><p>Simulating physics with computers</p></title>
  <aug>
    <au><snm>Feynman</snm><fnm>R</fnm></au>
  </aug>
  <source>International Journal of Theoretical Physics</source>
  <publisher>Kluwer Academic Publishers-Plenum Publishers</publisher>
  <pubdate>1982</pubdate>
  <volume>21</volume>
  <issue>6-7</issue>
  <fpage>467</fpage>
  <lpage>-488</lpage>
  <url>http://dx.doi.org/10.1007/bf02650179</url>
</bibl>

<bibl id="B6">
  <title><p>{Universal Quantum Simulators}</p></title>
  <aug>
    <au><snm>{Lloyd}</snm><fnm>S</fnm></au>
  </aug>
  <source>Science</source>
  <pubdate>1996</pubdate>
  <volume>273</volume>
  <issue>5278</issue>
  <fpage>1073</fpage>
  <lpage>1078</lpage>
</bibl>

<bibl id="B7">
  <title><p>Quantum supremacy using a programmable superconducting processor</p></title>
  <aug>
    <au><snm>Arute</snm><fnm>F</fnm></au>
    <au><snm>Arya</snm><fnm>K</fnm></au>
    <au><snm>Babbush</snm><fnm>R</fnm></au>
    <au><snm>Bacon</snm><fnm>D</fnm></au>
    <au><snm>Bardin</snm><fnm>JC</fnm></au>
    <au><snm>Barends</snm><fnm>R</fnm></au>
    <au><snm>Biswas</snm><fnm>R</fnm></au>
    <au><snm>Boixo</snm><fnm>S</fnm></au>
    <au><snm>Brandao</snm><fnm>FG</fnm></au>
    <au><snm>Buell</snm><fnm>DA</fnm></au>
    <au><cnm>others</cnm></au>
  </aug>
  <source>Nature</source>
  <publisher>Nature Publishing Group</publisher>
  <pubdate>2019</pubdate>
  <volume>574</volume>
  <issue>7779</issue>
  <fpage>505</fpage>
  <lpage>-510</lpage>
</bibl>

<bibl id="B8">
  <title><p>Noisy intermediate-scale quantum algorithms</p></title>
  <aug>
    <au><snm>Bharti</snm><fnm>K</fnm></au>
    <au><snm>Cervera Lierta</snm><fnm>A</fnm></au>
    <au><snm>Kyaw</snm><fnm>TH</fnm></au>
    <au><snm>Haug</snm><fnm>T</fnm></au>
    <au><snm>Alperin Lea</snm><fnm>S</fnm></au>
    <au><snm>Anand</snm><fnm>A</fnm></au>
    <au><snm>Degroote</snm><fnm>M</fnm></au>
    <au><snm>Heimonen</snm><fnm>H</fnm></au>
    <au><snm>Kottmann</snm><fnm>JS</fnm></au>
    <au><snm>Menke</snm><fnm>T</fnm></au>
    <au><cnm>others</cnm></au>
  </aug>
  <source>Reviews of Modern Physics</source>
  <publisher>APS</publisher>
  <pubdate>2022</pubdate>
  <volume>94</volume>
  <issue>1</issue>
  <fpage>015004</fpage>
</bibl>

<bibl id="B9">
  <title><p>Evidence for the utility of quantum computing before fault tolerance</p></title>
  <aug>
    <au><snm>Kim</snm><fnm>Y</fnm></au>
    <au><snm>Eddins</snm><fnm>A</fnm></au>
    <au><snm>Anand</snm><fnm>S</fnm></au>
    <au><snm>Wei</snm><fnm>KX</fnm></au>
    <au><snm>Van Den Berg</snm><fnm>E</fnm></au>
    <au><snm>Rosenblatt</snm><fnm>S</fnm></au>
    <au><snm>Nayfeh</snm><fnm>H</fnm></au>
    <au><snm>Wu</snm><fnm>Y</fnm></au>
    <au><snm>Zaletel</snm><fnm>M</fnm></au>
    <au><snm>Temme</snm><fnm>K</fnm></au>
    <au><cnm>others</cnm></au>
  </aug>
  <source>Nature</source>
  <publisher>Nature Publishing Group UK London</publisher>
  <pubdate>2023</pubdate>
  <volume>618</volume>
  <issue>7965</issue>
  <fpage>500</fpage>
  <lpage>-505</lpage>
</bibl>

<bibl id="B10">
  <title><p>Digital Quantum Simulation of Linear and Nonlinear Optical Elements</p></title>
  <aug>
    <au><snm>Sab{\'{\i}}n</snm><fnm>C</fnm></au>
  </aug>
  <source>Quantum Reports</source>
  <publisher>{MDPI} {AG}</publisher>
  <pubdate>2020</pubdate>
  <volume>2</volume>
  <issue>1</issue>
  <fpage>208</fpage>
  <lpage>-220</lpage>
  <url>https://doi.org/10.33902Fquantum2010013</url>
</bibl>

<bibl id="B11">
  <title><p>Quantum simulations of physics problems</p></title>
  <aug>
    <au><snm>Somma</snm><fnm>RD</fnm></au>
    <au><snm>Ortiz</snm><fnm>G</fnm></au>
    <au><snm>Knill</snm><fnm>EH</fnm></au>
    <au><snm>Gubernatis</snm><fnm>J</fnm></au>
  </aug>
  <source>{SPIE} Proceedings</source>
  <publisher>{SPIE}</publisher>
  <editor>Eric Donkor and Andrew R. Pirich and Howard E. Brandt</editor>
  <pubdate>2003</pubdate>
  <url>https://doi.org/10.11172F12.487249</url>
</bibl>

<bibl id="B12">
  <title><p>Quantum Computation, Complexity, and Many-Body Physics</p></title>
  <aug>
    <au><snm>Somma</snm><fnm>RD</fnm></au>
  </aug>
  <pubdate>2005</pubdate>
</bibl>

<bibl id="B13">
  <title><p>Digital quantum simulation of beam splitters and squeezing with IBM quantum computers</p></title>
  <aug>
    <au><snm>Encinar</snm><fnm>PC</fnm></au>
    <au><snm>Agust\'{\i}</snm><fnm>A</fnm></au>
    <au><snm>Sab\'{\i}n</snm><fnm>C</fnm></au>
  </aug>
  <source>Phys. Rev. A</source>
  <publisher>American Physical Society</publisher>
  <pubdate>2021</pubdate>
  <volume>104</volume>
  <fpage>052609</fpage>
  <url>https://link.aps.org/doi/10.1103/PhysRevA.104.052609</url>
</bibl>

<bibl id="B14">
  <title><p>Digital quantum simulation of quantum gravitational entanglement with {IBM} quantum computers</p></title>
  <aug>
    <au><snm>Sab{\'{\i}}n</snm><fnm>C</fnm></au>
  </aug>
  <source>{EPJ} Quantum Technology</source>
  <publisher>Springer Science and Business Media {LLC}</publisher>
  <pubdate>2023</pubdate>
  <volume>10</volume>
  <issue>1</issue>
  <url>https://doi.org/10.11402Fepjqt2Fs40507-023-00161-6</url>
</bibl>

<bibl id="B15">
  <title><p>Preparation of nonclassical states in cavities with a moving mirror</p></title>
  <aug>
    <au><snm>Bose</snm><fnm>S.</fnm></au>
    <au><snm>Jacobs</snm><fnm>K.</fnm></au>
    <au><snm>Knight</snm><fnm>P. L.</fnm></au>
  </aug>
  <source>Physical Review A</source>
  <publisher>American Physical Society ({APS})</publisher>
  <pubdate>1997</pubdate>
  <volume>56</volume>
  <issue>5</issue>
  <fpage>4175</fpage>
  <lpage>-4186</lpage>
  <url>https://doi.org/10.11032Fphysreva.56.4175</url>
</bibl>

<bibl id="B16">
  <title><p>Cavity optomechanics</p></title>
  <aug>
    <au><snm>Aspelmeyer</snm><fnm>M</fnm></au>
    <au><snm>Kippenberg</snm><fnm>TJ</fnm></au>
    <au><snm>Marquardt</snm><fnm>F</fnm></au>
  </aug>
  <source>Reviews of Modern Physics</source>
  <publisher>American Physical Society ({APS})</publisher>
  <pubdate>2014</pubdate>
  <volume>86</volume>
  <issue>4</issue>
  <fpage>1391</fpage>
  <lpage>-1452</lpage>
  <url>https://doi.org/10.11032Frevmodphys.86.1391</url>
</bibl>

<bibl id="B17">
  <title><p>An introduction to quantum optomechanics</p></title>
  <aug>
    <au><snm>Milburn</snm><fnm>GJ</fnm></au>
    <au><snm>Woolley</snm><fnm>MJ</fnm></au>
  </aug>
  <source>Acta Physica Slovaca</source>
  <pubdate>2011</pubdate>
  <volume>61</volume>
  <fpage>483</fpage>
  <lpage>601</lpage>
</bibl>

<bibl id="B18">
  <title><p>Quantum Computation and Quantum Information: 10th Anniversary Edition</p></title>
  <aug>
    <au><snm>Nielsen</snm><fnm>MA</fnm></au>
    <au><snm>Chuang</snm><fnm>IL</fnm></au>
  </aug>
  <publisher>Cambridge University Press</publisher>
  <pubdate>2011</pubdate>
</bibl>

<bibl id="B19">
  <title><p>Universal state inversion and concurrence in arbitrary dimensions</p></title>
  <aug>
    <au><snm>Rungta</snm><fnm>P</fnm></au>
    <au><snm>Bu{\v{z} }ek</snm><fnm>V.</fnm></au>
    <au><snm>Caves</snm><fnm>CM</fnm></au>
    <au><snm>Hillery</snm><fnm>M.</fnm></au>
    <au><snm>Milburn</snm><fnm>G. J.</fnm></au>
  </aug>
  <source>Physical Review A</source>
  <publisher>American Physical Society ({APS})</publisher>
  <pubdate>2001</pubdate>
  <volume>64</volume>
  <issue>4</issue>
  <url>https://doi.org/10.11032Fphysreva.64.042315</url>
</bibl>

<bibl id="B20">
  <title><p>Entanglement of a Pair of Quantum Bits</p></title>
  <aug>
    <au><snm>Hill</snm><fnm>S</fnm></au>
    <au><snm>Wootters</snm><fnm>WK</fnm></au>
  </aug>
  <source>Physical Review Letters</source>
  <publisher>American Physical Society ({APS})</publisher>
  <pubdate>1997</pubdate>
  <volume>78</volume>
  <issue>26</issue>
  <fpage>5022</fpage>
  <lpage>-5025</lpage>
  <url>https://doi.org/10.11032Fphysrevlett.78.5022</url>
</bibl>

<bibl id="B21">
  <title><p>Validating quantum computers using randomized model circuits</p></title>
  <aug>
    <au><snm>Cross</snm><fnm>AW</fnm></au>
    <au><snm>Bishop</snm><fnm>LS</fnm></au>
    <au><snm>Sheldon</snm><fnm>S</fnm></au>
    <au><snm>Nation</snm><fnm>PD</fnm></au>
    <au><snm>Gambetta</snm><fnm>JM</fnm></au>
  </aug>
  <source>Physical Review A</source>
  <publisher>American Physical Society ({APS})</publisher>
  <pubdate>2019</pubdate>
  <volume>100</volume>
  <issue>3</issue>
  <url>https://doi.org/10.11032Fphysreva.100.032328</url>
</bibl>

<bibl id="B22">
  <title><p>Probabilistic error cancellation for dynamic quantum circuits</p></title>
  <aug>
    <au><snm>Gupta</snm><fnm>RS</fnm></au>
    <au><snm>Berg</snm><fnm>E</fnm></au>
    <au><snm>Takita</snm><fnm>M</fnm></au>
    <au><snm>Riste</snm><fnm>D</fnm></au>
    <au><snm>Temme</snm><fnm>K</fnm></au>
    <au><snm>Kandala</snm><fnm>A</fnm></au>
  </aug>
  <pubdate>2023</pubdate>
</bibl>

</refgrp>
} 


\begin{thebibliography}{22}
\ifx \bisbn   \undefined \def \bisbn  #1{ISBN #1}\fi
\ifx \binits  \undefined \def \binits#1{#1}\fi
\ifx \bauthor  \undefined \def \bauthor#1{#1}\fi
\ifx \batitle  \undefined \def \batitle#1{#1}\fi
\ifx \bjtitle  \undefined \def \bjtitle#1{#1}\fi
\ifx \bvolume  \undefined \def \bvolume#1{\textbf{#1}}\fi
\ifx \byear  \undefined \def \byear#1{#1}\fi
\ifx \bissue  \undefined \def \bissue#1{#1}\fi
\ifx \bfpage  \undefined \def \bfpage#1{#1}\fi
\ifx \blpage  \undefined \def \blpage #1{#1}\fi
\ifx \burl  \undefined \def \burl#1{\textsf{#1}}\fi
\ifx \doiurl  \undefined \def \doiurl#1{\textsf{#1}}\fi
\ifx \betal  \undefined \def \betal{\textit{et al.}}\fi
\ifx \binstitute  \undefined \def \binstitute#1{#1}\fi
\ifx \binstitutionaled  \undefined \def \binstitutionaled#1{#1}\fi
\ifx \bctitle  \undefined \def \bctitle#1{#1}\fi
\ifx \beditor  \undefined \def \beditor#1{#1}\fi
\ifx \bpublisher  \undefined \def \bpublisher#1{#1}\fi
\ifx \bbtitle  \undefined \def \bbtitle#1{#1}\fi
\ifx \bedition  \undefined \def \bedition#1{#1}\fi
\ifx \bseriesno  \undefined \def \bseriesno#1{#1}\fi
\ifx \blocation  \undefined \def \blocation#1{#1}\fi
\ifx \bsertitle  \undefined \def \bsertitle#1{#1}\fi
\ifx \bsnm \undefined \def \bsnm#1{#1}\fi
\ifx \bsuffix \undefined \def \bsuffix#1{#1}\fi
\ifx \bparticle \undefined \def \bparticle#1{#1}\fi
\ifx \barticle \undefined \def \barticle#1{#1}\fi
\ifx \bconfdate \undefined \def \bconfdate #1{#1}\fi
\ifx \botherref \undefined \def \botherref #1{#1}\fi
\ifx \url \undefined \def \url#1{\textsf{#1}}\fi
\ifx \bchapter \undefined \def \bchapter#1{#1}\fi
\ifx \bbook \undefined \def \bbook#1{#1}\fi
\ifx \bcomment \undefined \def \bcomment#1{#1}\fi
\ifx \oauthor \undefined \def \oauthor#1{#1}\fi
\ifx \citeauthoryear \undefined \def \citeauthoryear#1{#1}\fi
\ifx \endbibitem  \undefined \def \endbibitem {}\fi
\ifx \bconflocation  \undefined \def \bconflocation#1{#1}\fi
\ifx \arxivurl  \undefined \def \arxivurl#1{\textsf{#1}}\fi
\csname PreBibitemsHook\endcsname

\bibitem{Bose_2017}
\begin{botherref}
\oauthor{\bsnm{Bose}, \binits{S.}},
\oauthor{\bsnm{Mazumdar}, \binits{A.}},
\oauthor{\bsnm{Morley}, \binits{G.W.}},
\oauthor{\bsnm{Ulbricht}, \binits{H.}},
\oauthor{\bsnm{Toro{\v{s} }}, \binits{M.}},
\oauthor{\bsnm{Paternostro}, \binits{M.}},
\oauthor{\bsnm{Geraci}, \binits{A.A.}},
\oauthor{\bsnm{Barker}, \binits{P.F.}},
\oauthor{\bsnm{Kim}, \binits{M.S.}},
\oauthor{\bsnm{Milburn}, \binits{G.J.}}:
Spin entanglement witness for quantum gravity.
Physical Review Letters
\textbf{119}(24)
(2017).
doi:\doiurl{10.1103/physrevlett.119.240401}
\end{botherref}
\endbibitem

\bibitem{Marletto_2017}
\begin{botherref}
\oauthor{\bsnm{Marletto}, \binits{C.}},
\oauthor{\bsnm{Vedral}, \binits{V.}}:
Gravitationally induced entanglement between two massive particles is sufficient evidence of quantum effects in gravity.
Physical Review Letters
\textbf{119}(24)
(2017).
doi:\doiurl{10.1103/physrevlett.119.240402}
\end{botherref}
\endbibitem

\bibitem{Bose}
\begin{barticle}
\bauthor{\bsnm{Bose}, \binits{S.}},
\bauthor{\bsnm{Mazumdar}, \binits{A.}},
\bauthor{\bsnm{Schut}, \binits{M.}},
\bauthor{\bparticle{Toro\ifmmode~\check{s}\else} \bsnm{\v{s}\fi{}}, \binits{M.}}:
\batitle{Mechanism for the quantum natured gravitons to entangle masses}.
\bjtitle{Phys. Rev. D}
\bvolume{105},
\bfpage{106028}
(\byear{2022}).
doi:\doiurl{10.1103/PhysRevD.105.106028}
\end{barticle}
\endbibitem

\bibitem{clarke}
\begin{botherref}
\oauthor{\bsnm{Clarke}, \binits{J.}},
\oauthor{\bsnm{Neveu}, \binits{P.}},
\oauthor{\bsnm{Khosla}, \binits{K.E.}},
\oauthor{\bsnm{Verhagen}, \binits{E.}},
\oauthor{\bsnm{Vanner}, \binits{M.R.}}:
Cavity quantum optomechanical nonlinearities and position measurement beyond the breakdown of the linearized approximation.
\arxivurl{2207.11153}
\end{botherref}
\endbibitem



\bibitem{biswas2023gravitational}
\begin{botherref}
\oauthor{\bsnm{Biswas}, \binits{D.}},
\oauthor{\bsnm{Bose}, \binits{S.}},
\oauthor{\bsnm{Mazumdar}, \binits{A.}},
\oauthor{\bsnm{Toroš}, \binits{M.}}:
Gravitational Optomechanics: Photon-Matter Entanglement via Graviton Exchange
(2023).
\arxivurl{2209.09273}
\end{botherref}
\endbibitem

\bibitem{Feynman1982Simulating}
\begin{barticle}
\bauthor{\bsnm{Feynman}, \binits{R.}}:
\batitle{Simulating physics with computers}.
\bjtitle{International Journal of Theoretical Physics}
\bvolume{21}(\bissue{6-7}),
\bfpage{467}--\blpage{488}
(\byear{1982}).
doi:\doiurl{10.1007/bf02650179}
\end{barticle}
\endbibitem

\bibitem{1996Sci...273.1073L}
\begin{barticle}
\bauthor{\bsnm{{Lloyd}}, \binits{S.}}:
\batitle{{Universal Quantum Simulators}}.
\bjtitle{Science}
\bvolume{273}(\bissue{5278}),
\bfpage{1073}--\blpage{1078}
(\byear{1996}).
doi:\doiurl{10.1126/science.273.5278.1073}
\end{barticle}
\endbibitem

\bibitem{arute2019quantum}
\begin{barticle}
\bauthor{\bsnm{Arute}, \binits{F.}},
\bauthor{\bsnm{Arya}, \binits{K.}},
\bauthor{\bsnm{Babbush}, \binits{R.}},
\bauthor{\bsnm{Bacon}, \binits{D.}},
\bauthor{\bsnm{Bardin}, \binits{J.C.}},
\bauthor{\bsnm{Barends}, \binits{R.}},
\bauthor{\bsnm{Biswas}, \binits{R.}},
\bauthor{\bsnm{Boixo}, \binits{S.}},
\bauthor{\bsnm{Brandao}, \binits{F.G.}},
\bauthor{\bsnm{Buell}, \binits{D.A.}}, \betal:
\batitle{Quantum supremacy using a programmable superconducting processor}.
\bjtitle{Nature}
\bvolume{574}(\bissue{7779}),
\bfpage{505}--\blpage{510}
(\byear{2019})
\end{barticle}
\endbibitem

\bibitem{bharti2022noisy}
\begin{barticle}
\bauthor{\bsnm{Bharti}, \binits{K.}},
\bauthor{\bsnm{Cervera-Lierta}, \binits{A.}},
\bauthor{\bsnm{Kyaw}, \binits{T.H.}},
\bauthor{\bsnm{Haug}, \binits{T.}},
\bauthor{\bsnm{Alperin-Lea}, \binits{S.}},
\bauthor{\bsnm{Anand}, \binits{A.}},
\bauthor{\bsnm{Degroote}, \binits{M.}},
\bauthor{\bsnm{Heimonen}, \binits{H.}},
\bauthor{\bsnm{Kottmann}, \binits{J.S.}},
\bauthor{\bsnm{Menke}, \binits{T.}}, \betal:
\batitle{Noisy intermediate-scale quantum algorithms}.
\bjtitle{Reviews of Modern Physics}
\bvolume{94}(\bissue{1}),
\bfpage{015004}
(\byear{2022})
\end{barticle}
\endbibitem

\bibitem{kim2023evidence}
\begin{barticle}
\bauthor{\bsnm{Kim}, \binits{Y.}},
\bauthor{\bsnm{Eddins}, \binits{A.}},
\bauthor{\bsnm{Anand}, \binits{S.}},
\bauthor{\bsnm{Wei}, \binits{K.X.}},
\bauthor{\bsnm{Van Den~Berg}, \binits{E.}},
\bauthor{\bsnm{Rosenblatt}, \binits{S.}},
\bauthor{\bsnm{Nayfeh}, \binits{H.}},
\bauthor{\bsnm{Wu}, \binits{Y.}},
\bauthor{\bsnm{Zaletel}, \binits{M.}},
\bauthor{\bsnm{Temme}, \binits{K.}}, \betal:
\batitle{Evidence for the utility of quantum computing before fault tolerance}.
\bjtitle{Nature}
\bvolume{618}(\bissue{7965}),
\bfpage{500}--\blpage{505}
(\byear{2023})
\end{barticle}
\endbibitem

\bibitem{Sab_n_2020}
\begin{barticle}
\bauthor{\bsnm{Sab{\'{\i}}n}, \binits{C.}}:
\batitle{Digital quantum simulation of linear and nonlinear optical elements}.
\bjtitle{Quantum Reports}
\bvolume{2}(\bissue{1}),
\bfpage{208}--\blpage{220}
(\byear{2020}).
doi:\doiurl{10.3390/quantum2010013}
\end{barticle}
\endbibitem

\bibitem{Somma_2003}
\begin{bchapter}
\bauthor{\bsnm{Somma}, \binits{R.D.}},
\bauthor{\bsnm{Ortiz}, \binits{G.}},
\bauthor{\bsnm{Knill}, \binits{E.H.}},
\bauthor{\bsnm{Gubernatis}, \binits{J.}}:
\bctitle{Quantum simulations of physics problems}.
In: \beditor{\bsnm{Donkor}, \binits{E.}},
\beditor{\bsnm{Pirich}, \binits{A.R.}},
\beditor{\bsnm{Brandt}, \binits{H.E.}} (eds.)
\bbtitle{{SPIE} Proceedings}.
\bpublisher{{SPIE}}
(\byear{2003}).
doi:\doiurl{10.1117/12.487249}.
\burl{https://doi.org/10.11172F12.487249}
\end{bchapter}
\endbibitem

\bibitem{somma2005quantum}
\begin{botherref}
\oauthor{\bsnm{Somma}, \binits{R.D.}}:
Quantum Computation, Complexity, and Many-Body Physics
(2005).
\arxivurl{quant-ph/0512209}
\end{botherref}
\endbibitem

\bibitem{PhysRevA.104.052609}
\begin{barticle}
\bauthor{\bsnm{Encinar}, \binits{P.C.}},
\bauthor{\bsnm{Agust\'{\i}}, \binits{A.}},
\bauthor{\bsnm{Sab\'{\i}n}, \binits{C.}}:
\batitle{Digital quantum simulation of beam splitters and squeezing with ibm quantum computers}.
\bjtitle{Phys. Rev. A}
\bvolume{104},
\bfpage{052609}
(\byear{2021}).
doi:\doiurl{10.1103/PhysRevA.104.052609}
\end{barticle}
\endbibitem

\bibitem{Sab_n_2023}
\begin{botherref}
\oauthor{\bsnm{Sab{\'{\i}}n}, \binits{C.}}:
Digital quantum simulation of quantum gravitational entanglement with {IBM} quantum computers.
{EPJ} Quantum Technology
\textbf{10}(1)
(2023).
doi:\doiurl{10.1140/epjqt/s40507-023-00161-6}
\end{botherref}
\endbibitem

\bibitem{opto58}
\begin{barticle}
\bauthor{\bsnm{Bose}, \binits{S.}},
\bauthor{\bsnm{Jacobs}, \binits{K.}},
\bauthor{\bsnm{Knight}, \binits{P.L.}}:
\batitle{Preparation of nonclassical states in cavities with a moving mirror}.
\bjtitle{Physical Review A}
\bvolume{56}(\bissue{5}),
\bfpage{4175}--\blpage{4186}
(\byear{1997}).
doi:\doiurl{10.1103/physreva.56.4175}
\end{barticle}
\endbibitem

\bibitem{opto59}
\begin{barticle}
\bauthor{\bsnm{Aspelmeyer}, \binits{M.}},
\bauthor{\bsnm{Kippenberg}, \binits{T.J.}},
\bauthor{\bsnm{Marquardt}, \binits{F.}}:
\batitle{Cavity optomechanics}.
\bjtitle{Reviews of Modern Physics}
\bvolume{86}(\bissue{4}),
\bfpage{1391}--\blpage{1452}
(\byear{2014}).
doi:\doiurl{10.1103/revmodphys.86.1391}
\end{barticle}
\endbibitem

\bibitem{optomechanics}
\begin{barticle}
\bauthor{\bsnm{Milburn}, \binits{G.J.}},
\bauthor{\bsnm{Woolley}, \binits{M.J.}}:
\batitle{An introduction to quantum optomechanics}.
\bjtitle{Acta Physica Slovaca}
\bvolume{61},
\bfpage{483}--\blpage{601}
(\byear{2011})
\end{barticle}
\endbibitem

\bibitem{Nielsen}
\begin{bbook}
\bauthor{\bsnm{Nielsen}, \binits{M.A.}},
\bauthor{\bsnm{Chuang}, \binits{I.L.}}:
\bbtitle{Quantum Computation and Quantum Information: 10th Anniversary Edition}.
\bpublisher{Cambridge University Press}
(\byear{2011})
\end{bbook}
\endbibitem

\bibitem{Rungta_2001}
\begin{botherref}
\oauthor{\bsnm{Rungta}, \binits{P.}},
\oauthor{\bsnm{Bu{\v{z} }ek}, \binits{V.}},
\oauthor{\bsnm{Caves}, \binits{C.M.}},
\oauthor{\bsnm{Hillery}, \binits{M.}},
\oauthor{\bsnm{Milburn}, \binits{G.J.}}:
Universal state inversion and concurrence in arbitrary dimensions.
Physical Review A
\textbf{64}(4)
(2001).
doi:\doiurl{10.1103/physreva.64.042315}
\end{botherref}
\endbibitem

\bibitem{Hill_1997}
\begin{barticle}
\bauthor{\bsnm{Hill}, \binits{S.}},
\bauthor{\bsnm{Wootters}, \binits{W.K.}}:
\batitle{Entanglement of a pair of quantum bits}.
\bjtitle{Physical Review Letters}
\bvolume{78}(\bissue{26}),
\bfpage{5022}--\blpage{5025}
(\byear{1997}).
doi:\doiurl{10.1103/physrevlett.78.5022}
\end{barticle}
\endbibitem

\bibitem{Cross_2019}
\begin{botherref}
\oauthor{\bsnm{Cross}, \binits{A.W.}},
\oauthor{\bsnm{Bishop}, \binits{L.S.}},
\oauthor{\bsnm{Sheldon}, \binits{S.}},
\oauthor{\bsnm{Nation}, \binits{P.D.}},
\oauthor{\bsnm{Gambetta}, \binits{J.M.}}:
Validating quantum computers using randomized model circuits.
Physical Review A
\textbf{100}(3)
(2019).
doi:\doiurl{10.1103/physreva.100.032328}
\end{botherref}
\endbibitem

\bibitem{gupta2023probabilistic}
\begin{botherref}
\oauthor{\bsnm{Gupta}, \binits{R.S.}},
\oauthor{\bparticle{van~den} \bsnm{Berg}, \binits{E.}},
\oauthor{\bsnm{Takita}, \binits{M.}},
\oauthor{\bsnm{Riste}, \binits{D.}},
\oauthor{\bsnm{Temme}, \binits{K.}},
\oauthor{\bsnm{Kandala}, \binits{A.}}:
Probabilistic error cancellation for dynamic quantum circuits
(2023).
\arxivurl{2310.07825}
\end{botherref}
\endbibitem

\end{thebibliography}





\section*{Tables}
\begin{table}[H]
\caption{Gate decomposition of each term in our Hamiltonian evolution operator.}
\centering
\begin{tabular}{|c|c|c|}
\hline
                  & $H_0$ & $U$ \\ \hline
$U_\varepsilon^{\left(1\right)}=e^{-i\tilde{\varepsilon}\sigma_x^0\sigma_x^1\sigma_x^2\sigma_x^3}$ & $-\sigma_z^0$      & $e^{i\frac{\pi}{4}\sigma_x^0}
e^{i\frac{\pi}{4}\sigma_z^0\sigma_x^1}e^{i\frac{\pi}{4}\sigma_z^0\sigma_x^2}e^{i\frac{\pi}{4}\sigma_z^0\sigma_x^3}$     \\ \hline
$U_\varepsilon^{\left(2\right)}=e^{-i\tilde{\varepsilon}\sigma_x^0\sigma_x^1\sigma_y^2\sigma_y^3}$ & $-\sigma_z^0$      &  $e^{i\frac{\pi}{4}\sigma_x^0}
e^{i\frac{\pi}{4}\sigma_z^0\sigma_x^1}e^{i\frac{\pi}{4}\sigma_z^0\sigma_x^2}e^{i\frac{\pi}{4}\sigma_z^0\sigma_x^3}e^{i\frac{\pi}{4}\sigma_z^2}e^{i\frac{\pi}{4}\sigma_z^3}$    \\ \hline
$U_\varepsilon^{\left(3\right)}=e^{-i\tilde{\varepsilon}\sigma_y^0\sigma_y^1\sigma_x^2\sigma_x^3}$ & $-\sigma_z^0$      &  $e^{i\frac{\pi}{4}\sigma_x^0}
e^{i\frac{\pi}{4}\sigma_z^0\sigma_x^1}e^{i\frac{\pi}{4}\sigma_z^0\sigma_x^2}e^{i\frac{\pi}{4}\sigma_z^0\sigma_x^3}e^{i\frac{\pi}{4}\sigma_z^0}e^{i\frac{\pi}{4}\sigma_z^1}$    \\ \hline
$U_\varepsilon^{\left(4\right)}=e^{-i\tilde{\varepsilon}\sigma_y^0\sigma_y^1\sigma_y^2\sigma_y^3}$ & $-\sigma_z^0$      & $e^{i\frac{\pi}{4}\sigma_x^0}
e^{i\frac{\pi}{4}\sigma_z^0\sigma_x^1}e^{i\frac{\pi}{4}\sigma_z^0\sigma_x^2}e^{i\frac{\pi}{4}\sigma_z^0\sigma_x^3}e^{i\frac{\pi}{4}\sigma_z^0}e^{i\frac{\pi}{4}\sigma_z^1}e^{i\frac{\pi}{4}\sigma_z^2}e^{i\frac{\pi}{4}\sigma_z^3}$     \\ \hline
\end{tabular}
\end{table}
\end{backmatter}
\end{document}